\documentclass[10pt,conference]{IEEEtran}\IEEEoverridecommandlockouts
\usepackage{epsfig,graphics,subfigure,psfrag,amsmath,cases}
\usepackage{latexsym,amssymb,amsmath,epsfig,subfigure,algorithm}
\usepackage{algorithmic,enumerate}
\usepackage{color}
\usepackage{url}
\usepackage{scrtime}
\usepackage{array}

\author{\authorblockN{ Derrick Wing Kwan Ng\authorrefmark{1}, Robert Schober\authorrefmark{2}, and Hussein Alnuweiri\authorrefmark{3}
\thanks{\authorrefmark{2}The author is also with the University of British Columbia, Vancouver, Canada. This research was supported by the Qatar National Research Fund (QNRF), under project NPRP 5-401-2-161.  }}
Institute for Digital Communications, Friedrich-Alexander-University Erlangen-N\"urnberg (FAU), Germany\authorrefmark{1}\authorrefmark{2}\\
Texas A\&M University at Qatar,  Qatar\authorrefmark{3}\\
Email: kwan@lnt.de,  schober@lnt.de, hussein.alnuweiri@qatar.tamu.edu\vspace*{-4mm}

}

\title{Secure Layered Transmission in Multicast Systems with Wireless Information and Power Transfer}

\date{\thistime,\,\today}

\newtheorem{Thm}{Theorem}
\newtheorem{Lem}{Lemma}
\newtheorem{Cor}{Corollory}

 \newcommand{\qed}{\hfill \ensuremath{\blacksquare}}
\DeclareMathOperator{\Tr}{Tr}
\DeclareMathOperator{\Rank}{\mathrm{Rank}}
\DeclareMathOperator{\card}{\mathrm{card}}

\DeclareMathOperator{\maxo}{\mathrm{maximize}}
\DeclareMathOperator{\mino}{\mathrm{minimize}}

\newcommand{\abs}[1]{\lvert#1\rvert}
\newcommand{\norm}[1]{\lVert#1\rVert}
\begin{document}

\maketitle

\begin{abstract}
This paper considers downlink
multicast transmit beamforming for secure layered transmission systems with wireless simultaneous information and power transfer.  We study the power allocation algorithm design for  minimizing the total transmit power in the presence of  passive eavesdroppers and energy harvesting receivers.  The algorithm design is formulated  as a non-convex optimization problem. Our problem formulation promotes the dual use of energy signals in providing secure communication and facilitating efficient energy transfer. Besides, we  take into account a minimum required  power for energy harvesting  at the idle receivers and heterogeneous quality of service (QoS) requirements for the multicast video receivers. In light of the intractability of the problem, we reformulate
the considered problem by replacing a non-convex probabilistic constraint with a convex deterministic constraint which leads to a smaller feasible solution set. Then, a semidefinite programming
 relaxation (SDR) approach is adopted to obtain an upper bound solution for the reformulated problem. Subsequently, sufficient conditions for the global optimal solution of the reformulated problem are revealed.  Furthermore,
we propose  two suboptimal power allocation schemes based on the upper bound solution. Simulation results demonstrate the excellent
performance and significant transmit power savings  achieved by the proposed  schemes compared to isotropic energy signal generation.

\end{abstract}
\renewcommand{\baselinestretch}{0.90}
\large\normalsize

\section{Introduction}
\label{sect1}
In recent years, using  multimedia applications over wireless communication channels such as internet protocol television (IPTV) and video streaming  has become increasingly popular. The associated high data  rate requirements  have
led to a  tremendous demand  for energy and bandwidth. As a result, advanced signal processing techniques for multiple antenna transmitters and physical layer multicasting have been proposed in the literature for facilitating power efficient  communication services \cite{CN:Multicast}\nocite{JR:Gaussian_randomization}--\cite{JR:TWC_large_antennas}. However, the   lifetime of networks remains the bottleneck for system performance. In particular,  mobile devices are often powered by  capacity-limited batteries and energy is dissipated even if the receivers are idle.

The integration of  energy harvesting technology with   communication devices  provides  self-sustainability  to power-constrained communication networks. Hydroelectric, piezoelectric, solar, and wind are the major conventional energy sources for energy harvesting. However, the availability of
these natural energy sources is usually  limited by location or
 climate and they may not be available in  indoor environments.
  On the other hand, wireless power transfer, where energy is harvested from ambient radio signals in radio frequency (RF), is also a viable source of energy for energy scavenging \cite{CN:WIPT_fundamental}--\nocite{CN:Shannon_meets_tesla,JR:MIMO_WIPT}\cite{JR:WIPT_fullpaper}. Furthermore,   wireless energy harvesting technology facilitates the possibility of simultaneous wireless information and power transfer. In \cite{CN:WIPT_fundamental} and \cite{CN:Shannon_meets_tesla}, the fundamental tradeoff between the harvested energy and the achievable data rate was investigated for flat fading and frequency selective channels, respectively.  In \cite{JR:MIMO_WIPT}, the authors studied the design of precoders for multiple antenna systems to achieve various information and energy transmission tradeoffs.   In \cite{JR:WIPT_fullpaper}, the authors proposed   power allocation schemes for maximizing the energy efficiency of systems with concurrent information and power transfer.  Yet,  the results in \cite{CN:WIPT_fundamental}--\cite{JR:WIPT_fullpaper} are obtained by assuming  single layer transmission  which does not capture the properties of multilayer transmission  in multimedia applications.

Layered transmission  is a promising approach for achieving a better resource utilization in multimedia applications and has been implemented in different multimedia standards such as JPEG2000 and H.264/SVC \cite{JR:picutre_layers,JR:Video_layers3}.   Thereby, for instance, a video  source signal is encoded into  multiple layers with different source coding rates, i.e.,
 a base layer and several enhancement layers.  In particular, the base layer can be  decoded independently from the other layers and the embedded information provides a
basic video quality; the enhancement layers can only
be decoded successively together with the base layer and they further refine
the quality of the video information encoded in the base layer. As a result,  the structure of layered transmission
requires  unequal error protection which  introduces a paradigm shift in resource allocation algorithm design. On the other hand, a large amount of work has been  recently devoted
to  physical (PHY) layer security
\cite{Report:Wire_tap}--\nocite{JR:Artifical_Noise1,JR:Kwan_physical_layer}\cite{JR:Kwan_secure_imperfect},
  as a complement to traditional cryptographic encryption adopted in the application layer. Specifically,  PHY layer security guarantees   perfectly secure communication by exploiting the physical characteristics of the wireless  communication channel.  In \cite{Report:Wire_tap}, Wyner showed that when the legitimate receiver enjoys a better channel quality than the eavesdropper, the transmitter
 can deliver  perfectly secure messages to the legitimate receiver at
a non-zero data rate. In \cite{JR:Artifical_Noise1} and \cite{JR:Kwan_physical_layer}, artificial noise generation was proposed for providing secure communication in fast fading and slow fading channels, respectively.   In \cite{JR:Kwan_secure_imperfect},   beamforming design was proposed to minimize the total transmit power of the system with simultaneous  energy and secure information transfer. However, single layer transmission was adopted and the receivers in \cite{Report:Wire_tap}--\cite{JR:Kwan_secure_imperfect} were assumed to be powered by perpetual energy sources which may not be valid for power-constrained portable devices requiring multimedia services.  Furthermore,  an efficient power allocation scheme for secure layered transmission systems with  energy harvesting receivers has not been reported in the literature \cite{CN:Multicast}--\cite{JR:Kwan_secure_imperfect}.

 \begin{figure*}[t]
 \centering\vspace*{-3mm}
\includegraphics[width=4.3in]{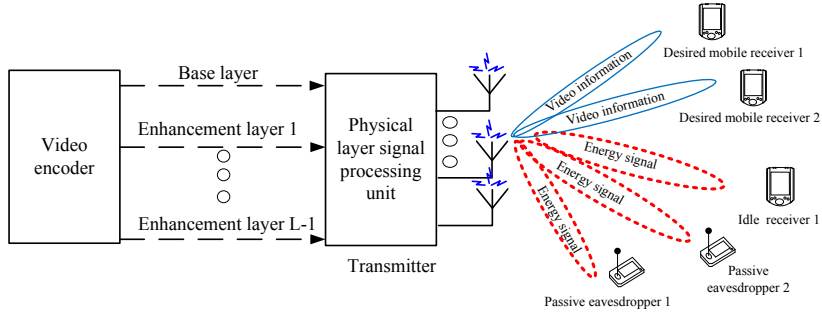}
 \caption{A single group multicast communication system with layered transmission for $K=3$ mobile receivers (two active and one idle) and $J=2$ passive eavesdroppers. } \label{fig:system_model}\vspace*{-3mm}
\end{figure*}

In this paper, we address the above issues and study the power allocation
 algorithm design for multicast secure layered  transmission systems with simultaneous  wireless information and power transfer.  The algorithm design is formulated  as a non-convex optimization problem. To circumvent the problem intractability, we reformulate the optimization problem by introducing a convex tractable deterministic constraint. Subsequently, a semidefinite programming relaxation (SDR)  based power allocation algorithm is proposed to obtain an upper bound solution for the reformulated problem. Besides, the upper bound solution is used as a building block for the design of two suboptimal schemes which are shown to achieve excellent system performance.

\section{System Model}
\label{sect:OFDMA_AF_network_model}
In this section,  we  present the adopted system model for secure layered video transmission.
\subsection{Channel  Model}
We consider a single group\footnote{The results of this paper can be generalized to multigroup multicast networks at the expense of a more involved notation.} multicast downlink communication  which comprises   a transmitter with $N_\mathrm{T}$ transmit antennas, $J$ single antenna passive eavesdroppers,   and $K$  single antenna video receivers,  cf. Figure \ref{fig:system_model}. In every time instant, the  transmitter conveys video information to a  given set of multicast video service  subscribers (receivers) requiring  the same video information  while the remaining  receivers are idle.  The eavesdroppers are assumed to be passive and silent to hide their existences.
 We focus on a time division duplexing system with a frequency flat slow time varying fading channel. The
downlink channel gains of all legitimate receivers can be accurately obtained at the transmitter via measuring  the uplink pilot sequences in the handshaking signals by exploiting channel reciprocity. The downlink
received signals at desired video receiver $k\in\{1,\ldots,K\}$ and passive eavesdropper $j\in\{1,\ldots,J\}$ are given by, respectively,
\begin{eqnarray}
y_k&=&\mathbf{h}^{H}_k \mathbf{x}+z_{\mathrm{s},k}\\
y_{\mathrm{PE},j}&=&\mathbf{g}_{j}^{H} \mathbf{x} +z_{\mathrm{e},j},
\end{eqnarray}
where $\mathbf{x}\in\mathbb{C}^{ N_\mathrm{T} \times 1}$ denotes the transmitted symbol vector and $\mathbb{C}^{N\times M}$ denotes the space of $N\times M$ matrices with complex entries.
$\mathbf{h}^{H}_k\in\mathbb{C}^{1\times N_\mathrm{T}}$ is the channel
vector between the transmitter and  legitimate receiver $k$
  and
$\mathbf{g}_{j}^{H} \in\mathbb{C}^{1\times N_\mathrm{T}}$ is the channel
vector between the transmitter and passive eavesdropper $j$. $(\cdot)^H$ denotes the conjugate transpose of a matrix.
  $z_{\mathrm{s},k}$ and $z_{\mathrm{e},j}$ denote the joint effect of
  thermal noise and signal processing noise at  desired video receiver $k$ and passive eavesdropper $j$, respectively. They are modeled as  additive white Gaussian noises   with zero mean and variance $\sigma_{\mathrm{s}}^2$.

\subsection{Information Decoding and Energy Harvesting Receiver}
\label{sect:receiver}
We assume that each video receiver has the ability to  harvest energy and to decode the modulated information\footnote{Please refer to \cite{CN:Shannon_meets_tesla} for a hardware implementation of
 the energy harvesting circuitry.} from the received radio signal. However, in practice,
 the signal used for decoding of the modulated
information cannot be reused for harvesting energy due to hardware limitations \cite{JR:MIMO_WIPT}. As a result,   a video receiver can either decode the video information  when it is active (being served by the transmitter) or harvest energy when it is idle,
 but not performing both simultaneously, cf. Figure \ref{fig:system_model}. Besides, a rechargeable battery is used to store the harvested energy for future use. If the energy storage of the battery is full, the energy harvesting process will stop to prevent the battery from being overcharged.    To facilitate energy harvesting at the idle receivers, the transmitter can increase the transmit power of the video information signal. However, this makes the video information more vulnerable to eavesdropping due to the high transmit power.  As a result,  a novel power allocation algorithm design is needed for promoting simultaneous  power and  information  transfer while ensuring robustness against passive eavesdropping.

\subsection{Video Encoding and Energy Signal Generation}
A  layered video encoding approach is adopted to encode the video information. Specifically,
 the video source signal is encoded  into $L$ layers at the transmitter and  the data rate of each layer is fixed, cf. H.264/SVC  \cite{JR:Video_layers,JR:Video_layers2}. The video information can be represented  as $\mathbf{S}=\big[s_1,s_2,\ldots,s_L\big],
s_l\in\mathbb{C},\forall l\in \{1,\ldots,L\}$, where $s_l$ denotes the video information of layer $l$.  The $L$ layers include one base layer which
 contains the most essential information of the video. The base layer can be decoded independently without utilizing information from the other layers and the embedded information  guarantees a minimum video  quality. The remaining $L-1$ layers are  enhancement layers which are used to successively refine the  decoded lower layers. However, the information contained in a given enhancement layer cannot be decoded  if there is a failure in decoding any of the lower layers.

On the other
hand,  an energy
signal is transmitted along with the information signals to degrade the quality of the channels of the eavesdroppers and to facilitate  energy harvesting at the idle receivers. Hence, the transmitter adopts the transmit symbol vector $\mathbf{x}$ as
\begin{eqnarray}
\mathbf{x}=\hspace*{-2mm}\underbrace{\sum_{l=1}^L \mathbf{w}_l s_l}_{\mbox{$L$-layer video signal}}+\underbrace{\mathbf{w}_{\mathrm{E}}}_{\mbox{energy signal}},
\end{eqnarray}
where $\mathbf{w}_l\in\mathbb{C}^{N_\mathrm{T}\times 1}$ is the beamforming vector for the video information in layer $l$ intended for the desired receivers. We note that  superposition coding is used to superimpose the $L$  video information layers. $\mathbf{w}_{\mathrm{E}}\in\mathbb{C}^{N_\mathrm{T}\times 1}$ is the energy signal vector generated to cause interference to the passive eavesdroppers. In particular,  $\mathbf{w}_{\mathrm{E}}$ is modeled as  a complex Gaussian pseudo-random vector represented as
 $\mathbf{w}_{\mathrm{E}}\sim {\cal CN}(\mathbf{0}, \mathbf{W}_{\mathrm{E}})$. $\mathbf{W}_{\mathrm{E}}$ denotes the covariance matrix of the energy signal, i.e., $\mathbf{W}_{\mathrm{E}}\in \mathbb{H}^{N_\mathrm{T}}$ and $\mathbf{W}_{\mathrm{E}}\succeq \mathbf{0}$. Here, $\mathbf{W}_{\mathrm{E}}\succeq\mathbf{0}$ indicates that $ \mathbf{W}_{\mathrm{E}}$ is a  positive semidefinite Hermitian matrix and $\mathbb{H}^N$ represents the set of all $N$-by-$N$ complex Hermitian matrices. Besides, the  Gaussian pseudo-random energy signal is known at the  legitimate receivers but is unavailable at the passive eavesdroppers. The transmitter regularly  changes the seeds of the random sequence generator used for generating the energy signal sequences to prevent the sequence from being cracked by the passive eavesdroppers. Besides, the seeds information used at the transmitter can be delivered to  the desired receivers securely  by exploiting e.g. the reciprocity of the channels between the transmitter and the desired receivers \cite{CN:secrect_key}.

\section{Power Allocation Algorithm Design}\label{sect:forumlation}
In this section, we present the adopted performance metrics and the problem formulation.

\subsection{Channel Capacity}
\label{subsect:Instaneous_Mutual_information}
We assume that perfect channel state information is available at the video receivers. Besides,  successive interference cancellation \cite{book:david_wirelss_com} is performed at the receivers decoding the video information. Since the energy signal sequence is known at the legitimate receivers,  the energy signal, $\mathbf{w}_{\mathrm{E}}$, is first removed from the received signal before video information decoding starts. Then, the  receivers decode and cancel the lower video information layers  successively  before decoding the higher layers. Therefore,  the instantaneous channel capacity between the transmitter and legitimate mobile video receiver $k\in\{1,\ldots,K\}$ in layer $l\in\{1,\ldots,L\}$
is given by
\begin{eqnarray}
C_{l,k}&=&\log_2\Big(1+\Gamma_{l,k}\Big)\,\,\,\,
\mbox{and}\,\,\\
\Gamma_{l,k}&=&\frac{\abs{\mathbf{h}^H_k\mathbf{w}_l}^2}
{\sum_{t=l+1}^L\abs{\mathbf{h}^H_k\mathbf{w}_t}^2+\sigma_\mathrm{s}^2}\label{eqn:cap} ,
\end{eqnarray}
where $\Gamma_{l,k}$ is the received signal-to-interference-plus-noise ratio (SINR) of layer $l$ at  receiver $k$, and   $\abs{\cdot}$  denotes   the absolute value of a complex scalar.

On the other hand, the channel state information of the passive eavesdroppers is not available at the transmitter.  As a result, we design the power allocation algorithm for the worst case scenario. In particular,   we assume that
eavesdropper $j$ is close to the transmitter and is located at the reference distance of the path loss model. Then,  an upper bound on the capacity of layer $l$  between the transmitter and passive eavesdropper $j\in\{1,\ldots,J\}$ under this worst case scenario is given by \cite{JR:Artifical_Noise1,JR:Kwan_physical_layer}
\begin{eqnarray}  \label{eqn:eavesdropper-SINR-bound}
C_{\mathrm{PE}_{l,j}}\hspace*{-3mm}&=&\hspace*{-3mm}\log_2\Big(1+\Gamma_{\mathrm{PE}_{l,j}}\Big)\,\,\,\,
\mbox{and}\,\,\\
\Gamma_{\mathrm{PE}_{l,j}\hspace*{-3mm}}&=&\hspace*{-3mm}\frac{\mathbf{w}^H_l\mathbf{G}_j\mathbf{w}_l}{
\sum_{t=l+1}^L\abs{\mathbf{g}^H_k\mathbf{w}_t}^2+\Tr(\mathbf{G}_j\mathbf{W}_{\mathrm{E}})+\sigma_\mathrm{s}^2 }\notag\\ \label{eqn:SINR_up_passive}
\hspace*{-3mm}&\stackrel{(a)}{\le} &\hspace*{-3mm} \frac{\mathbf{w}^H_l\mathbf{{\tilde G}}_j\mathbf{w}_l}
{\sum_{t=l+1}^L\mathbf{w}^H_t\mathbf{\tilde G}_j\mathbf{w}_t+\Tr(\mathbf{\tilde G}_j\mathbf{W}_{\mathrm{E}})+\tilde\sigma_{\mathrm{s},j}^2},
\end{eqnarray}
where $\Tr(\cdot)$ denotes the trace of a matrix, $\mathbf{G}_j=\mathbf{g}_j\mathbf{g}_j^H$, $\tilde\sigma_{\mathrm{s},j}^2=\frac{\sigma_\mathrm{s}^2 }{{\cal E}\{\norm{\mathbf{g}_j^{\mathrm{Ref}}}^2\}}$, and  $\mathbf{g}_j^{\mathrm{Ref}}$ contains the channel coefficients of the eavesdropper at the reference distance.  Here, $\norm{\cdot}$ and  ${\cal E}\{\cdot\}$  represent the Euclidean vector norm and statistical expectation, respectively. We note that the elements of $\mathbf{g}_j^{\mathrm{Ref}}$ and $\mathbf{g}_j$  capture the joint effect of small scale fading and shadowing in the same manner. Yet, the path loss coefficient contained in $\mathbf{g}_j^{\mathrm{Ref}}$ is calculated at the reference distance which results in $\norm{\mathbf{g}_j^{\mathrm{Ref}}}\ge\norm{\mathbf{g}_j}$. Since ${\cal E}\{\norm{\mathbf{g}_1^{\mathrm{Ref}}}^2\}=\ldots={\cal E}\{\norm{\mathbf{g}_j^{\mathrm{Ref}}}^2\}=\ldots={\cal E}\{\norm{\mathbf{g}_J^{\mathrm{Ref}}}^2\}$, we replace $\tilde\sigma_{\mathrm{s},j}^2$ by $\tilde\sigma_{\mathrm{s}}^2$ without loss of generality. Besides, $\mathbf{{\tilde G}}_j=\mathbf{\tilde g}_j\mathbf{\tilde g}_j^H=\frac{\mathbf{g}_j^{\mathrm{Ref}}(\mathbf{g}_j^{\mathrm{Ref}})^H}{{\cal E}\{\norm{\mathbf{g}_j^{\mathrm{Ref}}}^2\}}$ in $(a)$ is a normalized matrix with $\norm{\mathbf{\tilde g}_j}=1$. Furthermore, the passive eavesdroppers are unable to perform interference cancellation to remove $\Tr(\mathbf{G}_j\mathbf{W}_\mathrm{E})$ since the energy signal $\mathbf{w}_{\mathrm{E}}$ is only known at the legitimate receivers.   With a slight abuse of notation, we reuse variable  $C_{\mathrm{PE}_{l,j}}$  to
 denote the  upper bound on the  capacity  of layer $l$ at passive eavesdropper $j$ by replacing the SINR $\Gamma_{\mathrm{PE}_{l,j}}$  in (\ref{eqn:eavesdropper-SINR-bound}) with its upper bound in (\ref{eqn:SINR_up_passive}). Thus, the maximum secrecy capacity of layer $l$ between the transmitter
and the desired active receivers is given by \cite{JR:Artifical_Noise1}
\begin{eqnarray}\label{eqn:secrecy_cap}
C_{\mathrm{sec}_l}=\Big[\min_{k\in {\cal A}}C_{l,k} - \underset{j\in\{1,\ldots,J\}}{\max} C_{\mathrm{PE}_{l,j}}\Big]^+,
\end{eqnarray}
where  $[x]^+=\max\{0,x\}$ and $\cal A$ denotes the set of active receivers which  require video information.
\subsection{Optimization Problem Formulation}
\label{sect:cross-Layer_formulation}
The optimal power allocation policy $\{\mathbf{w}_l^*,\mathbf{W}^*_{\mathrm{E}}\}$ can be obtained by solving:
\begin{eqnarray}
\label{eqn:cross-layer}&&\hspace*{0mm} \underset{\mathbf{W}_{\mathrm{E}}\in \mathbb{H}^{N_\mathrm{T}},\mathbf{w}_l
}{\mino} \,\, \sum_{l=1}^L\norm{\mathbf{w}_l}^2+\Tr(\mathbf{W}_{\mathrm{E}})\nonumber\\
\notag \mbox{s.t.} \hspace*{-1mm}&&\hspace*{-5mm}\mbox{C1: }\notag\frac{\abs{\mathbf{h}^H_k\mathbf{w}_l}^2}{\overset{L}{\underset{t=l+1}{\sum}} \abs{\mathbf{h}^H_k\mathbf{w}_t}^2+\sigma_{\mathrm{s}}^2} \ge \Gamma_{\mathrm{req}_l}, \forall k\in {\cal P},\forall l, \\
\hspace*{-1mm}&&\hspace*{-5mm}\mbox{C2: }\notag\frac{\abs{\mathbf{h}_k^H\mathbf{w}_1}^2}
{\overset{L}{\underset{t=2}{\sum}} \abs{\mathbf{h}_k^H\mathbf{w}_t}^2+\sigma_{\mathrm{s}}^2} \ge \Gamma_{\mathrm{req}_1},\forall k\in {\cal B},\\
\hspace*{-1mm}&&\hspace*{-5mm}\mbox{C3: }\notag\Pr\Big(\max_{j\in\{1,\ldots,J\}} \Big\{\Gamma_{\mathrm{PE}_{1,j}}\Big\} \le \Gamma_{\mathrm{tol}_1}\Big)\ge \kappa, \\
\hspace*{-1mm}&&\hspace*{-5mm}\mbox{C4: }\notag \sum_{l=1}^L\abs{\mathbf{h}_k^H\mathbf{w}_l}^2+\Tr(\mathbf{H}_k\mathbf{W}_{\mathrm{E}})\ge\frac{ P_{\min_k}}{\eta_k},\forall  k\in {\cal I},\\
\hspace*{-1mm}&&\hspace*{-5mm}\mbox{C5:}\,\, \mathbf{W}_{\mathrm{E}}\succeq \mathbf{0},
\end{eqnarray}
where $\mathbf{H}_k=\mathbf{h}_k\mathbf{h}_k^H$. $\Gamma_{\mathrm{req}_l}$ in C1 is the minimum required SINR  for decoding layer $l$ at receiver $k$ and  $\cal P$ is the set of receivers which  subscribe the  \emph{premium video service}. In particular, the transmitter is required to  guarantee the quality of service (QoS) (i.e., SINR) of each layer for the \emph{premium video service}. $\cal B$ in C2  denotes the set of receivers which subscribe the \emph{basic video service}. Constraint C2 indicates that the transmitter only guarantees a minimum required SINR for the first layer which provides the basic video quality.   In C3, $\Gamma_{\mathrm{tol}_1}$ denotes the  maximum receive SINR  tolerance in layer 1 for decoding layer 1 successfully  at passive eavesdropper $j$. In particular, the maximum SINR  among all passive eavesdroppers is required to be smaller than  $\Gamma_{\mathrm{tol}_1}$ with at least probability $\kappa$.  Since layered coding is employed for  video information encoding,    it is sufficient to protect the  first  video information layer against passive eavesdropping. If an eavesdropper is unable to decode layer 1, then he/she will also not be able to decode layer $l\ge2$.   Besides, although  the number of eavesdroppers  $J$ is not known at the transmitter, $J$ in C3 represents the maximum tolerable number of passive eavesdroppers that the transmitter can handle. Furthermore, we do not  maximize the secrecy capacity of video delivery in this paper as it does not necessarily lead to a power efficient power allocation. Yet, the problem formulation in (\ref{eqn:cross-layer})
guarantees  a minimum secrecy capacity of layer 1, i.e., $C_{\mathrm{sec}_1}\ge
\log_2(1+\Gamma_{\mathrm{req}_1})-\log_2(1+\Gamma_{\mathrm{tol}_1})$, with probability $\kappa$.  Moreover, $\cal I$ in C4 represents the set of idle receivers and  $P_{\min_k}$ denotes the minimum required   power  harvested at idle receiver $k$.  $0 <\eta_k \le 1$ is a constant
which denotes the efficiency of the energy harvesting circuit for
converting the received radio signal to electrical energy. We note that  sets $\cal P, I, B$ are mutually exclusive and $\cal P,  B\in \cal A$. C5
 and $\mathbf{W}_{\mathrm{E}}\in \mathbb{H}^{N_\mathrm{T}}$  are imposed such that $\mathbf{W}_{\mathrm{E}}$ satisfies the requirements on the covariance matrix of the energy signal. We also point out that the benefits of energy signal generation are two-fold. First, the energy signal is used to compromise the channels of the passive eavesdroppers for providing communication security, cf. C3 and (\ref{eqn:SINR_up_passive}). Second, it acts as a energy source for the idle receivers for energy harvesting, cf. C4.
\section{Solution of the Optimization Problem} \label{sect:solution}
The problem  in (\ref{eqn:cross-layer}) is a  non-convex optimization problem. In order to obtain a tractable power allocation algorithm, we first handle constraints C1, C2, and C4 by transforming the problem into an equivalent. Subsequently,  we reformulate the considered problem by replacing the probabilistic constraint C3 with  a convex deterministic constraint. Then, an  semidefinite programming relaxation (SDR) approach is adopted to obtain an upper bound solution for the reformulated problem. Finally, we propose two suboptimal power allocation schemes which achieve excellent system performances.
\subsection{Semidefinite Programming Relaxation} \label{sect:solution_dual_decomposition}
First, we  rewrite problem (\ref{eqn:cross-layer})  in an equivalent form via SDP:
\begin{eqnarray}
\label{eqn:SDP}&&\hspace*{-2mm} \underset{\mathbf{W}_l,\mathbf{W}_{\mathrm{E}}\in \mathbb{H}^{N_\mathrm{T}}
}{\mino}\,\, \sum_{l=1}^L \Tr(\mathbf{W}_l)+\Tr(\mathbf{W}_{\mathrm{E}})\nonumber\\
\notag \mbox{s.t.} \hspace*{-1mm}&&\hspace*{-5mm}\mbox{C1: }\notag\frac{\Tr(\mathbf{H}_k\mathbf{W}_l)}{ \hspace*{-3mm} \overset{L}{\underset{t=l+1}{\sum}}\hspace*{-0.5mm} \Tr(\mathbf{H}_k\mathbf{W}_t)\hspace*{-0.5mm}+\hspace*{-0.5mm}\sigma_{\mathrm{s}}^2}\hspace*{-0.5mm} \ge\hspace*{-0.5mm} \Gamma_{\mathrm{req}_l}, \forall k\in \hspace*{-0.5mm}{\cal P},\forall l, \\
\hspace*{-1mm}&&\hspace*{-5mm}\mbox{C2: }\notag\frac{\Tr(\mathbf{H}_k\mathbf{W}_1)}
{\overset{L}{\underset{t=2}{\sum}}\hspace*{-0.5mm} \Tr(\mathbf{H}_k\mathbf{W}_t)\hspace*{-0.5mm}+\hspace*{-0.5mm}\sigma_{\mathrm{s}}^2} \hspace*{-0.5mm}\ge\hspace*{-0.5mm} \Gamma_{\mathrm{req}_1},\forall k\in\hspace*{-0.5mm} {\cal B},\\
\hspace*{-1mm}&&\hspace*{-5mm}\mbox{C3: }\notag\Pr\Big(\max_{j\in\{1,\ldots,J\}} \Big\{\Gamma_{\mathrm{PE}_{1,j}}\Big\} \le \Gamma_{\mathrm{tol}_1}\Big)\ge \kappa, \\
\hspace*{-1mm}&&\hspace*{-5mm}\mbox{C4: }\notag \Tr\Big(\mathbf{H}_k (\mathbf{W}_{\mathrm{E}}+\sum_{l=1}^L\mathbf{W}_l)\Big)\ge \frac{P_{\min_k}}{\eta_k},\forall  k\in {\cal I},\\
\hspace*{-1mm}&&\hspace*{-5mm}\mbox{C5:}\,\, \mathbf{W}_{\mathrm{E}}\succeq \mathbf{0},\quad\mbox{C6:}\,\, \mathbf{W}_l\succeq \mathbf{0},\,\,\forall l\in\{1,\ldots,L\},\notag\\
\hspace*{-1mm}&&\hspace*{-5mm}\mbox{C7:}\,\, \Rank(\mathbf{W}_l)=1,\,\,\forall l\in\{1,\ldots,L\},
\end{eqnarray}
where   $\mathbf{W}_l=\mathbf{w}_l\mathbf{w}^H_l$ and $\Rank(\cdot)$ in C7 denotes the rank of an input matrix. We note that $\mathbf{W}_l\succeq \mathbf{0},\forall l\in\{1,\ldots,L\}$, $\mathbf{W}_l\in \mathbb{H}^{N_\mathrm{T}},\forall l$, and $\Rank(\mathbf{W}_l)=1,\forall l,$ in (\ref{eqn:SDP}) are imposed to guarantee that $\mathbf{W}_l=\mathbf{w}_l\mathbf{w}^H_l$ holds after optimizing $\mathbf{W}_l$.  The transformed problem above is still non-convex due to the probabilistic  constraint C3 and the rank constraint in C7.  To overcome this problem, we introduce the following lemma for reformulating the considered problem:
  \begin{Lem}\label{lemma:chance_constraint} Assuming the normalized upper bound channel gain vectors of the passive eavesdroppers can be modeled as independent and identical distributed (i.i.d.) Rayleigh random variables,  $\mathbf{\tilde g}_j$, the following constraint implication holds:
\begin{eqnarray}\label{eqn:chance_constraint}\notag
&&\hspace*{-6mm} \overline{\mbox{C3}}\mbox{: }\Phi^{-1}_{N_{\mathrm{T}}}(1-\kappa^{1/J})\Gamma_{\mathrm{tol}_1}\tilde\sigma^2_{\mathrm{s}}\ge \lambda_{\max}\Big(\mathbf{Q}\Big)\\
\Rightarrow&&\hspace*{-6mm}{\mbox{C3: }}\Pr\Big(\max_{j\in\{1,\ldots,J\}} \Big\{\Gamma_{\mathrm{PE}_{1,j}}\Big\} \le \Gamma_{\mathrm{tol}_1}\Big)\ge \kappa,
\end{eqnarray}
\end{Lem}
where $\mathbf{Q}=\mathbf{W}_1\hspace*{-1mm}-\hspace*{-1mm}\Gamma_{\mathrm{tol}_1}(\sum_{t=2}^L\mathbf{W}_t\hspace*{-1mm}+\hspace*{-1mm}\mathbf{W}_{\mathrm{E}})$, $\Phi^{-1}_{N_{\mathrm{T}}}(\cdot)$ denotes the inverse
cumulative distribution function (c.d.f.) of an inverse central
chi-square random variable with 2$N_\mathrm{T}$ degrees of freedom and $\lambda_{\max}(\cdot)$ denotes the maximum eigenvalue of a  square input matrix.

 \emph{\,Proof:} Please refer to Appendix A.\qed

As a result, we can replace the probabilistic  constraint C3 with the constraint in (\ref{eqn:chance_constraint}). Constraint $\overline{\mbox{C3}}$ is safe and tractable in the sense that: (i) a feasible solution point satisfying  $\overline{\mbox{C3}}$ will also satisfy C3, and (ii) the new constraint  $\overline{\mbox{C3}}$ is  a convex function with respect to the optimization variables\footnote{Although the maximum eigenvalue function is a non-smooth function, it is a convex function \cite{book:convex}. }.

 Thus,  replacing probabilistic  constraint C3 with $\overline{\mbox{C3}}$ in (\ref{eqn:chance_constraint}), we obtain the following reformulated optimization problem:
\begin{eqnarray}
\label{eqn:SDP-robust}&&\hspace*{-5mm} \underset{\mathbf{W}_l,\mathbf{W}_{\mathrm{E}}\in \mathbb{H}^{N_{\mathrm{T}}}
}{\mino}\,\,  \sum_{l=1}^L \Tr(\mathbf{W}_l)+\Tr(\mathbf{W}_{\mathrm{E}})\\
\notag \mbox{s.t.} &&\hspace*{10mm}\mbox{C1, C2, C4, C5, C6, C7,} \\
&&\hspace*{-8mm}\overline{\mbox{C3}}\mbox{: }\notag \Gamma_{\mathrm{tol}_1}\tilde\sigma_{\mathrm{s}}^2\Phi^{-1}_{N_{\mathrm{T}}}(1-\kappa^{1/J})\ge \lambda_{\max}\Big(\mathbf{Q})\Big).
\end{eqnarray}
We note that the solution of the reformulated problem in (\ref{eqn:SDP-robust}) serves as a performance lower bound of (\ref{eqn:SDP}) as a smaller feasible solution set is considered by replacing probabilistic  constraint C3 with $\overline{\mbox{C3}}$. Now, C7: $\Rank(\mathbf{W}_l)=1,\forall l$, is the remaining obstacle in solving the reformulated problem in (\ref{eqn:SDP-robust}).  In fact, if this constraint is removed from the problem formulation, the reformulated problem becomes a convex SDP which satisfies Slater's constraint qualification. Thus,  the SDP relaxed version of the reformulated problem in (\ref{eqn:SDP-robust}) can be solved efficiently by off-the-shelf numerical solvers such as SeDuMi \cite{JR:SeDumi}.  In particular, if the obtained solution admits a rank-one matrix $\mathbf{W}_l,\forall l$, then it is the optimal solution of the original problem in (\ref{eqn:SDP-robust}). Yet, the proposed constraint relaxation may not be tight with respect to the reformulated problem and the result of the relaxed problem serves as a performance upper bound for the reformulated problem. Therefore, it is important to study under what condition(s) the solution of the problem in (\ref{eqn:SDP-robust}) yields rank-one matrices $\mathbf{W}_l,\forall l$.

 \subsection{Optimality Conditions}
We investigate  some sufficient conditions for obtaining a rank-one solution $\mathbf{W}_l$ via the  Karush-Kuhn-Tucker (KKT) conditions  of problem (\ref{eqn:SDP-robust}). To this end, we express
the Lagrangian function  of  (\ref{eqn:SDP-robust}) as
\begin{eqnarray}\notag\hspace*{-2mm}&&{\cal
L}(\mathbf{W}_l,\mathbf{W}_{\mathrm{E}},\boldsymbol \mu,\boldsymbol{\beta},\boldsymbol{\delta},\phi,\mathbf{Y}_l,\mathbf{X})\\
\notag\hspace*{-5mm}&=&\hspace*{-3mm} \sum_{l=1}^L \Tr(\mathbf{W}_l(\mathbf{I}_{N_{\mathrm{T}}}-\mathbf{Y}_l))- \phi\Phi^{-1}_{N_{\mathrm{T}}}(1-\kappa^{1/J})\Gamma_{\mathrm{tol}_1}\tilde\sigma_{\mathrm{s}}^2\notag\\
\notag\hspace*{-5mm}&-&\hspace*{-3mm}\Tr(\mathbf{W}_{\mathrm{E}}\mathbf{X})\hspace*{-0.2mm}+\hspace*{-0.2mm}\sum_{k\in \cal P}\sum_{l=1}^L
\mu_{l,k}\Big(\hspace*{-1mm}\frac{-\Tr(\mathbf{H}_k\mathbf{W}_l)}{\Gamma_{\mathrm{req}_l}}\hspace*{-0.5mm}+\hspace*{-0.5mm}\sum_{t=l+1}^L\hspace*{-1mm} \Tr(\mathbf{H}_k\mathbf{W}_l)\notag\Big.\\
\hspace*{-5mm}& +&\hspace*{-3mm}\sigma_{\mathrm{s}}^2\Big)+\phi \lambda_{\max}\Big(\mathbf{W}_1-\hspace*{-1mm}\Gamma_{\mathrm{tol}_1}(\sum_{t=2}^L\mathbf{W}_t+ \mathbf{W}_{\mathrm{E}})\Big)+\Tr(\mathbf{W}_{\mathrm{E}}) \notag\\
\notag\hspace*{-5mm}&+&\hspace*{-3mm} \sum_{k\in \cal B}\beta_k \Big(-\frac{\Tr(\mathbf{H}_k\mathbf{W}_1)}{\Gamma_{\mathrm{req}_1}}+\sum_{j=2}^L \Tr(\mathbf{H}_k\mathbf{W}_j)+\sigma_{\mathrm{s}}^2\Big)\\
\hspace*{-5mm}&+&\hspace*{-3mm}\sum_{k\in {\cal I}}\delta_k\Big(\frac{ P_{\min_k}}{ \eta_k} - \Tr\Big(\mathbf{H}_k( \mathbf{W}_{\mathrm{E}}+\sum_{l=1}^L \mathbf{W}_l )\Big)\Big),
\label{eqn:Lagrangian}
\end{eqnarray}
where $\mathbf{I}_{N_{\mathrm{T}}}$ is the $N_{\mathrm{T}}\times N_{\mathrm{T}}$ identity matrix.
$\boldsymbol\mu$, with elements $\mu_{l,k}\ge 0, \forall l\in\{1,\ldots,L\},\forall k\in {\cal P}$, is the Lagrange multiplier vector associated with  the minimum required SINR in decoding layer $l$ for receiver  $k$ subscribing  \emph{premium video service} in C1. $\boldsymbol \beta$, with elements $\beta_k\ge 0,\forall k\in {\cal B}$, is
the minimum required SINR in decoding layer $1$ for receiver $k$ requesting \emph{basic video service} in C2. $\phi$ is the Lagrange multiplier associated  with the proposed constraint $\overline{\mbox{C3}}$ from Lemma 1.  $\boldsymbol \delta$,  with elements $\delta_{k}\ge 0,\forall k\in {\cal I}$,  is the Lagrange multiplier vector
 for the minimum required  power transfer to the idle receivers in C4.  Matrices $\mathbf{X},\mathbf{Y}_l\succeq \mathbf{0}$ are the Lagrange multipliers for the positive semidefinite constraints on matrices $\mathbf{W}_{\mathrm{E}}$ and $\mathbf{W}_l$  in C5 and C6, respectively.  The resulting dual problem for the SDP relaxed problem is given by
\begin{eqnarray}\label{eqn:dual}
\underset{ \underset{\mathbf{Y}_l,\mathbf{X}\succeq \mathbf{0}}{\boldsymbol \mu,\boldsymbol{\beta},\boldsymbol{\delta},\phi\succeq \mathbf{0}}}{\maxo}\ \underset{{\mathbf{W}_l,\mathbf{W}_{\mathrm{E}}\in \mathbb{H}^{N_\mathrm{T}}}}{\mino}\,\,{\cal
L}(\mathbf{W}_l,\mathbf{W}_{\mathrm{E}},\boldsymbol \mu,\boldsymbol{\beta},\boldsymbol{\delta},\phi,\mathbf{Y}_l,\mathbf{X}).\label{eqn:master_problem}
\end{eqnarray}
In the following theorem, we reveal the rank of $\mathbf{W}_l^*$ in the relaxed version of problem (\ref{eqn:SDP-robust}).
\begin{Thm}In general, for $\Gamma_{\mathrm{req}_l}>0,\forall l,$ the following rank inequality for $\mathbf{W}_1^*$  holds:
\begin{eqnarray}
\Rank(\mathbf{W}^*_1)&\le& \min\{K,{N}_{\mathrm{T}}\}.
\end{eqnarray}
 Moreover, for $\phi=0$, i.e., constraint $\overline{\mbox{C3}}$ is not active at the optimal solution, there exist $\mathbf{W}^*_l$ such that
\begin{eqnarray}\notag
&&\sum_{l=1}^L\Rank^2(\mathbf{W}^*_l)+\Rank^2(\mathbf{W}^*_{\mathrm{E}})\notag\\
\le&& \card({\cal P}) L+\card({\cal B})+\card({\cal I})
\end{eqnarray}
\end{Thm}
where $\card(\cdot)$ denotes the cardinality of a set.

 \emph{\,Proof:} Please refer to Appendix B.\qed

By utilizing Theorem 1, we summarize some sufficient conditions for obtaining a rank-one solution via SDP relaxation in the following corollary.
\begin{Cor} For $\Gamma_{\mathrm{req}_l}>0,\forall l$, and $\phi=0$, the SDP relaxation of (\ref{eqn:SDP-robust}) is tight (i.e., $\Rank(\mathbf{W}_l)=1,\forall l$,) if one of the following is satisfied:
 \begin{enumerate}[i)]
 \item $L$ is arbitrary, $K\le 2:$ $\card({\cal P})=1$
    \item $ L=1, K=3: \card({\cal P})=2, \card({\cal B})=1,\card({\cal I})=0$
     \item $ L=1, K=3: \card({\cal P})=2, \card({\cal B})=0,\card({\cal I})=1$
 \end{enumerate}
\end{Cor}
We note that Corollary 1 provides only sufficient conditions for the optimality of SDP relaxation. In practice, the SDP relaxation can admit a rank-one solution   even if the sufficient conditions stated in Corollary 1 are not satisfied.

In the following, two suboptimal
power allocation schemes are designed based on the framework of convex optimization.

\subsubsection{Suboptimal Power Allocation Scheme 1}
The first proposed suboptimal power allocation scheme is a hybrid scheme based on the  solution of the relaxed version of (\ref{eqn:SDP-robust}). In particular, we first solve (\ref{eqn:SDP-robust}) by SDP relaxation. If the solution admits  rank-one $\mathbf{W}_l,\forall l$, then  the global optimal solution of  (\ref{eqn:SDP-robust}) is obtained. Otherwise, we  construct a suboptimal solution set $ \mathbf{\widetilde W}_l = \mathbf {\widetilde w}_l \mathbf {\widetilde w}_l^H$, where $\mathbf {\widetilde w}_l $ is the eigenvector corresponding to the maximum eigenvalue of matrix $\mathbf{W}_l$, where $\mathbf{W}_l$ is the solution of the SDP relaxed version of (\ref{eqn:SDP-robust}) with $\Rank(\mathbf{W}_l)>1$.  Then, we define $L$ scaling constants $\alpha_l,\forall l\in\{1,\ldots,L \}$
 and a new optimization problem:
\begin{eqnarray}\label{eqn:suboptimal1}\notag
&& \hspace*{-4mm}\underset{\alpha_l,\mathbf{W}_{\mathrm{E}}\in \mathbb{H}^{N_\mathrm{T}}
}{\mino}\,\, \sum_{l=1}^L \alpha_l\Tr(\mathbf{\widetilde W}_l)+\Tr(\mathbf{W}_{\mathrm{E}})\\
\notag \mbox{s.t.}
&&\hspace*{-5mm}\mbox{C1:}\notag\frac{\alpha_l\Tr(\mathbf{H}_k\mathbf{\widetilde W}_l)}{ \Tr\Big(\mathbf{H}_k(\mathbf{W}_{\mathrm{E}}\hspace*{-0.5mm}+\hspace*{-0.5mm} \overset{L}{\underset{t=l+1}{\sum}}\hspace*{-0.5mm}\alpha_t \mathbf{\widetilde W}_t
)\Big)\hspace*{-0.5mm}+\hspace*{-0.5mm}\sigma_{\mathrm{s}}^2}\hspace*{-0.5mm} \ge\hspace*{-0.5mm} \Gamma_{\mathrm{req}_l}, \forall k\in {\cal P},\forall l, \\
\hspace*{-1mm}&&\hspace*{-5mm}\mbox{C2:}\notag\frac{\alpha_1\Tr(\mathbf{H}_k\mathbf{W}_1)}
{ \Tr\Big( \mathbf{H}_k\big(\mathbf{W}_{\mathrm{E}}\hspace*{-0.5mm}+\hspace*{-0.5mm}\overset{L}{\underset{t=2}{\sum}}\hspace*{-0.5mm} \alpha_t\mathbf{\widetilde W}_t)\Big)\hspace*{-0.5mm}+\hspace*{-0.5mm}\sigma_{\mathrm{s}}^2} \hspace*{-0.5mm}\ge\hspace*{-0.5mm} \Gamma_{\mathrm{req}_1},\forall k\in {\cal B},\\
\hspace*{-1mm}&&\hspace*{-5mm}\overline{\mbox{C3}}\mbox{: }\Phi^{-1}_{N_{\mathrm{T}}}(1-\kappa^{1/J})\Gamma_{\mathrm{tol}_1}\tilde\sigma^2_{\mathrm{s}}\hspace*{-0.5mm}\ge \hspace*{-0.5mm} \lambda_{\max}\big(\mathbf{\tilde Q}\big),\notag \\
\hspace*{-1mm}&&\hspace*{-5mm}\mbox{C4: }\notag \Tr\Big(\mathbf{H}_k(\mathbf{W}_{\mathrm{E}}+ \sum_{l=1}^L\alpha_l \mathbf{\widetilde W}_l)\Big)\ge \frac{P_{\min_k}}{\eta_k},\forall  k\in {\cal I},\\
\hspace*{-1mm}&&\hspace*{-5mm}\mbox{C5:}\,\, \mathbf{W}_{\mathrm{E}}\succeq \mathbf{0},\quad\mbox{C6:}\,\, \alpha_l\ge 0,\forall l,
\end{eqnarray}
where $\mathbf{\tilde Q}=\alpha_1\mathbf{\widetilde  W}_1\hspace*{-1mm}-\hspace*{-1mm}\Gamma_{\mathrm{tol}_1}(\sum_{t=2}^L\alpha_t\mathbf{\widetilde W}_t+ \mathbf{W}_{\mathrm{E}})$. The problem formulation in (\ref{eqn:suboptimal1}) is convex with respect to the optimization variables. In particular, it  serves as a suboptimal solution for (\ref{eqn:SDP-robust}).

\subsubsection{Suboptimal Power  Allocation Scheme 2}
The second proposed suboptimal power allocation scheme is also a hybrid scheme. In particular, it is based  on the  solution of the relaxed version of (\ref{eqn:SDP-robust}) and
the rank-one Gaussian randomization scheme  \cite{JR:Gaussian_randomization}. Besides, a similar approach to solving the problem is adopted as in  suboptimal power  allocation scheme 1, except for the choice of beamforming matrix $\mathbf{\widetilde W}_l$ when  $\Rank(\mathbf{W}_l)>1$. Specifically, we calculate the eigenvalue decomposition of $\mathbf{W}_l=\mathbf{U}_l\mathbf{\Sigma}_l\mathbf{U}^H_l$, where $\mathbf{U}_l$ and $\mathbf{\Sigma}_l$  are an $N_\mathrm{T}\times N_\mathrm{T}$ unitary matrix and a diagonal matrix, respectively. Then, we adopt the suboptimal beamforming vector as $\mathbf {\widetilde w}_l=\mathbf{U}_l\mathbf{\Sigma}^{1/2}_l\mathbf{r}_l, \mathbf {\widetilde W}_l=\alpha_l\mathbf {\widetilde w}_l\mathbf {\widetilde w}_l^H$, where $\mathbf{r}_l\in {\mathbb C}^{N_{\mathrm{T}}}$ and $\mathbf{r}_l\sim {\cal CN}(\mathbf{0}, \mathbf{I}_{N_{\mathrm{T}}})$. Subsequently, we follow the same approach as in (\ref{eqn:suboptimal1}) for optimizing $\{\alpha_l,\mathbf{W}_{\mathrm{E}}\}$ and obtain a suboptimal rank-one solution $\alpha_l \mathbf{\widetilde W}_l$.

%

\section{Results}
\label{sect:result-discussion} In this section, we evaluate the
system performance for the proposed power allocation schemes using simulations.  We adopt the TGn path loss model \cite{report:tgn} for indoor communication with a reference distance of $2$ meters for the path loss model and  a carrier center frequency of $470$ MHz \cite{report:80211af}.  There are $K$ legitimate video receivers uniformly distributed between
the reference distance and the maximum service distance of $20$ meters.  The transmitter is equipped with $N_{\mathrm{T}}$ antennas and  we assume a joint transmit and receive antenna gain of  $10$ dBi (isotropic). The multipath fading coefficients between the transmitter and legitimate video receivers are generated as i.i.d. Rician random
variables with Rician factor $6$ dB.  On the other hand, we assume that  there are $J=4$ eavesdroppers eavesdropping the video information from outdoor\footnote{Although the eavesdroppers are located  outdoor, this location information is not known at the transmitter. The problem formulation in (\ref{eqn:cross-layer}) assumes Rayleigh fading channels for passive eavesdroppers and  considers the worst case scenario in (\ref{eqn:SINR_up_passive}) for providing secure communication. }. Thus,  the multipath fading coefficients between the transmitter and the $J$ passive eavesdroppers  are modeled as Rayleigh random variables.  The noise power and the RF energy conversion efficiency at the receivers are $\sigma_{\mathrm{s}}^2=-33$ dBm and $\eta_k=0.5,\forall k$, respectively. The video signal is encoded into 3 layers with minimum SINR requirements of $\Gamma_{\mathrm{req}_1}=6$ dB, $\Gamma_{\mathrm{req}_2}=9$ dB, and $\Gamma_{\mathrm{req}_3}=12$ dB, respectively. The maximum tolerable SINR at eavesdropper $j$ is set to $\Gamma_{\mathrm{tol}_1}=-10$ dB and $\kappa=0.99$. In other words, the  proposed power allocation schemes guarantee a minimum  secrecy capacity of layer 1 video information of  $C_{\mathrm{sec}_1}\ge  2.179$ bit/s/Hz with $0.99$ probability. Unless specified otherwise, we assume that there are two idle receivers requiring minimum harvested  powers of $P_{\min_k}=0$ dBm, $\forall k\in {\cal I}$, and two video receivers requiring basic video service. The system performance is obtained by averaging over $50000$ multipath  fading and path loss realizations.
\begin{figure}[t]
 \centering\vspace*{-3mm}
\includegraphics[width=3.5 in]{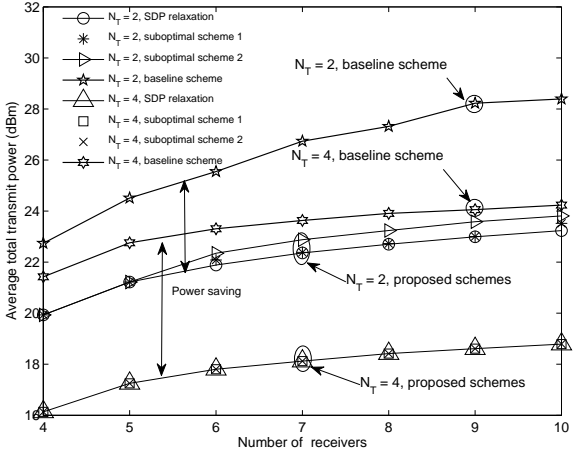}
\caption{Average total transmit power (dBm) versus the number of receivers for different power allocation schemes and different numbers of transmit antennas $N_{\mathrm{T}}$. The double-sided arrows indicate the power savings achieved by the proposed schemes compared to the baseline scheme.} \label{fig:p_SNR}\vspace*{-3mm}
\end{figure}

\subsection{Average Total Transmit Power }
Figure \ref{fig:p_SNR} depicts the  average total transmit power versus the number of video receivers $K$ for different power allocation schemes and different numbers of transmit antennas $N_{\mathrm{T}}$. It can be observed that the average total transmit power of the proposed schemes is a monotonically non-decreasing function of the number of video receivers. This is because a higher transmit power is required for satisfying constraint C1 when there are more video receivers requesting the premium video service. Besides, the two proposed suboptimal schemes perform closely to the upper bound system performance achieved by SDP relaxation. In fact,  the proposed suboptimal schemes 1 and 2 exploit the possibility of achieving the global optimal solution via SDP relaxation. On the other hand, the total transmit power decreases with an increasing number of transmit antennas. This is attributed to the fact that the degrees of freedom for power allocation increase with increasing number of transmit antennas which facilities the video information transmission.

For comparison, Figure
\ref{fig:p_SNR} also contains the average total transmit power of a baseline
power allocation scheme. The baseline scheme is a hybrid scheme. In particular, we adopt maximum ratio transmission (MRT) \cite{JR:TWC_large_antennas} for delivering  the video information of each layer with respect to the active receiver with the highest channel gain, i.e., $\max_{k\in {\cal A}} \norm{\mathbf{h}_k}^2$.  As for the energy signal, we adopt an isotropic radiation pattern for $\mathbf{W}_{\mathrm{E}}$. Then, we optimize both the power allocated to $\mathbf{W}_{\mathrm{E}}$ and the MRT beamforming vector for minimizing the total transmit power subject to the same constraints as in (\ref{eqn:SDP-robust}). Although the baseline scheme requires a lower computational complexity than the other schemes, it can be observed from  Figure
\ref{fig:p_SNR}  that the baseline scheme has the worst performance among all considered schemes. This is because  the transmitter is unable to fully exploit  the degrees of freedom in power allocation
when $\mathbf{W}_{\mathrm{E}}$ is radiated  isotropically and  $\mathbf{W}_l$ is fixed.

\subsection{Average Harvested Power}
\begin{figure}[t]
 \centering\vspace*{-3mm}
\includegraphics[width=3.5 in]{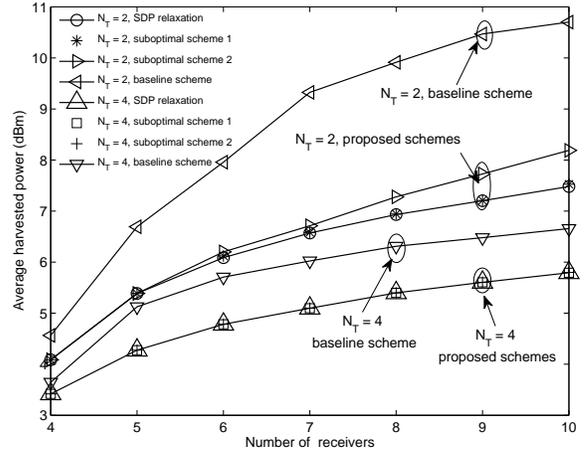}
\caption{Average total harvested power (dBm) versus the number of receivers for different power allocation schemes and different numbers of transmit antennas $N_{\mathrm{T}}$. } \label{fig:harvested_PT}\vspace*{-3mm}
\end{figure}

Figure \ref{fig:harvested_PT} shows the average harvested power versus the
number of receivers for different power allocation schemes. The average harvested power
increases with the number of receivers. In fact,  a higher transmit power is required for fulfilling the SINR requirements in C1 as the number of receivers increase which results in a higher energy level in the RF. Besides, the two proposed suboptimal power allocation schemes exhibit a similar performance as the SDR upper bound solution and are able to guarantee the minimum required harvesting power of the idle video receivers.   On the other hand,  as expected, the idle receivers are able to harvest more energy  for the baseline scheme than for the proposed schemes since for the baseline scheme, the transmitter has to transmit an exceedingly large amount of power in the RF to satisfy all the QoS requirements which benefits  energy harvesting.

\section{Conclusions}\label{sect:conclusion}
In this paper,  we  studied  the power allocation
algorithm design for secure layered  multicast  video transmission systems with simultaneous information and power transfer. The algorithm design was formulated as a non-convex optimization problem by taking into account energy signal generation to facilitate secure communication and efficient energy harvesting.  Due to the intractability of the problem, the
considered problem was reformulated by introducing a convex deterministic constraint. Subsequently, an SDP relaxation based power allocation algorithm was proposed to solve the  non-convex optimization problem which resulted in an upper bound solution of the reformulated problem. Besides, two suboptimal power allocation schemes were designed  exploiting the structure of the upper bound solution. Simulation results demonstrated the excellent performance of the proposed suboptimal schemes. In our future work, we will study the tightness of the proposed deterministic  constraint and  the impact of imperfect knowledge of the channel state information of the video receivers.

\section*{Appendix}

\subsection{Proof of Lemma 1}\label{sect:proof-lemma1}
By exploiting the independence of the $J$ passive eavesdroppers channels and after some mathematical manipulation, the left hand side of constraint C3 in (\ref{eqn:cross-layer}) can be equivalently written as
\begin{eqnarray}\label{eqn:C3-equivalent}
&&\Pr\Big(\Gamma_{\mathrm{tol}_1}\tilde\sigma_{\mathrm{s}}^2\ge\Tr(\mathbf{\tilde G}\mathbf{Q}) \Big)\ge \kappa^{1/J}.
\end{eqnarray}
Note that  we drop the index $j$ of the passive eavesdropper channels $\mathbf{\tilde G}_j$ in (\ref{eqn:C3-equivalent}) since the equivalent channels of the passive eavesdroppers are modeled as i.i.d. random variables.
 On the other hand,
the probability distribution in (\ref{eqn:C3-equivalent})  may not be a convex function and thus we focus on a smaller convex feasible solution set which provides a tractable solution. First, by utilizing the trace inequality in in \cite[Lemma II.1]{JR:trace_inequality}, we obtain
\begin{eqnarray} \label{eqn:apply-trace-inequality}
\Tr(\mathbf{\tilde G}\mathbf{Q})
\le\Tr(\mathbf{\tilde G})
\lambda_{\max}(\mathbf{Q}).
\end{eqnarray}
Then, by combining (\ref{eqn:C3-equivalent}) and (\ref{eqn:apply-trace-inequality}), 
the following implication holds:
\begin{eqnarray}\notag \label{eqn:implication}
&&\hspace*{-6mm}\Phi^{-1}_{N_{\mathrm{T}}}(1-\kappa^{1/J})\Gamma_{\mathrm{tol}_1}\tilde\sigma_{\mathrm{s}}^2\le \lambda_{\max}\big(\mathbf{Q}\big)\notag\notag\\
\hspace*{-6mm}\Longleftrightarrow&&\hspace*{-6mm}\Pr\Big(\frac{\lambda_{\max}(\mathbf{Q})}{\Gamma_{\mathrm{tol}_1}\tilde\sigma_{\mathrm{s}}^2}\le \frac{1}{\Tr(\mathbf{\tilde G})}\Big)\ge \kappa^{1/J}
\Rightarrow\mbox{(\ref{eqn:C3-equivalent})} \Rightarrow \mbox{C3},
\end{eqnarray}
where $\Phi^{-1}_{N_{\mathrm{T}}}(\cdot)$ denotes the inverse
cumulative distribution function (c.d.f.) of an inverse central
chi-square random variable with 2$N_\mathrm{T}$ degrees of freedom. We note that random variable $\mathbf{\tilde G}$ is decoupled from the  optimization variables, cf.  (\ref{eqn:apply-trace-inequality}). Thus,  the implication in (\ref{eqn:implication}) is applicable to any continuous channel distribution by replacing $\Phi^{-1}_{N_{\mathrm{T}}}(\cdot)$ with the inverse c.d.f. of  the corresponding distribution. On the other hand, the inverse c.d.f. can be obtained via a look-up table or by using the bisection method for practical implementation.

\subsection{Proof of Theorem 1}
 In the following, we study the rank of  the optimal $\mathbf{W}_l^*$ via focusing  on the corresponding KKT conditions:
\begin{eqnarray}\label{eqn:KKT}
\hspace*{-3mm}\mathbf{Y}^*_l\hspace*{-3mm}&\succeq&\hspace*{-3mm} \mathbf{0},\quad\beta_k^*,\,\delta_n^*,\,\mu_{l,k}^*,\,\phi^*\ge 0, \forall l,n,k,\\
\hspace*{-3mm} \mathbf{Y}^*_l\mathbf{W}^*_l\hspace*{-3mm}&=&\hspace*{-3mm}\mathbf{0},\forall l, \label{eqn:KKT-complementarity}\\
\mathbf{0}&\in &\partial_{\mathbf{W}_l} {\cal L}(\mathbf{\Xi}),\forall l, \label{eqn:subgradient}
\end{eqnarray}
where $\mathbf{\Xi}=\{\mathbf{W}_l,\mathbf{W}_{\mathrm{E}},\boldsymbol \mu,\boldsymbol{\beta},\boldsymbol{\delta},\phi,\mathbf{Y}_l,\mathbf{X}\}$ and $\mathbf{Y}^*_l,\,\beta_k^*,\,\delta_n^*,\,\mu_{l,k}^*,\,\phi^*$ are the optimal Lagrange multipliers for (\ref{eqn:dual}). Equation (\ref{eqn:KKT-complementarity}) is the complementary slackness condition. We note that $\mathbf{0}\in \partial_{\mathbf{W}_l} {\cal L}(\mathbf{\Xi})$ in (\ref{eqn:subgradient}) represents a set of sub-differentials with respect to $\mathbf{W}_l$   due to the non-differentiability of the maximum eigenvalue function in $\overline{\mbox{C3}}$. To obtain an expression for (\ref{eqn:subgradient}), we first define the sub-differential of the maximum eigenvalue function. The sub-differential of $\lambda_{\max}\big(\mathbf{Q}\big)$ with respect to $\mathbf{Q}$ can be expressed as \cite{JR:sub-diff-maximum-eigen}
\begin{eqnarray}\label{eqn:convex_hull_max_eigenvalue}
&&\partial_{\mathbf{Q}}\lambda_{\max}\big(\mathbf{Q}\big)
\\=&&\mathrm{convhull}\Big\{\mathbf{z}_n\mathbf{z}_n^H \,\,\Big| \norm{\mathbf{z}_n}^2=1, \mathbf{Q}\mathbf{z}_n=\lambda_{\max}\big(\mathbf{Q}\big)\mathbf{z}_n\Big\},\notag
\end{eqnarray}
where $\mathbf{z}_n$ is the $n$-th eigenvector of matrix $\mathbf{Q}$. Indeed, the sub-differential of the maximum eigenvalue function is a convex hull of the sub-gradients which satisfy  (\ref{eqn:convex_hull_max_eigenvalue}). Thus, without loss of generally, we can represent the element in the set which  satisfies $\mathbf{0}=\partial_{\mathbf{W}_1} {\cal L}(\mathbf{\Xi})$ as
\begin{eqnarray}\label{eqn:conve_hull_function}
\sum_{n=1}^{N_{\mathrm{T}}} \omega_n \mathbf{z}_n\mathbf{z}_n^H, \quad\mbox{where }\sum_{n=1}^{N_{\mathrm{T}}}\omega_n =1, \omega_n\ge 0, \omega_n\in\mathbb{R},\forall n,
\end{eqnarray}
and $\mathbb{R}$ denotes the set of real numbers. As a result, we can rewrite $\mathbf{0} \in\partial_{\mathbf{W}_1} {\cal L}(\mathbf{\Xi})$ as $\mathbf{0} =\partial_{\mathbf{W}_1} {\cal L}(\mathbf{\Xi})$ by applying (\ref{eqn:conve_hull_function}) to $\partial_{\mathbf{W}_1} {\cal L}(\mathbf{\Xi})$ which can be expressed as
 \begin{eqnarray}\label{eqn:post}
\label{eqn:KKT_Y1}
\hspace*{-3mm}\mathbf{Y}^*_1\hspace*{-3mm}&=&\hspace*{-3mm}\mathbf{I}_{N_\mathrm{T}}+\phi^*\sum_{n=1}^{N_{\mathrm{T}}} \omega_n \mathbf{z}_n\mathbf{z}_n^H\hspace*{-0.5mm}-\hspace*{-0.5mm}  \sum_{k \in{\cal P}} \frac{\mu_{1,k}^*\mathbf{H}_k}{\Gamma_{\mathrm{req}_1}} \notag\\
 &&\hspace*{-6mm}-  \sum_{k\in {\cal B}} \frac{\beta_k^*\mathbf{H}_k}{\Gamma_{\mathrm{req}_1}} -\sum_{k\in {\cal I}}\delta_k^* \mathbf{H}_k.
 \end{eqnarray}
 Then, we post-multiply both sides of (\ref{eqn:KKT_Y1}) by $\mathbf{W}^*_1$ and after exploiting (\ref{eqn:KKT-complementarity}) we obtain
 \begin{eqnarray}\label{eqn:post}
&&\hspace*{-3mm}\Big(\mathbf{I}_{N_\mathrm{T}}+\phi^*\sum_{n=1}^{N_{\mathrm{T}}} \omega_n \mathbf{z}_n\mathbf{z}_n^H \Big)\mathbf{W}^*_1 \\
=&&\hspace*{-3mm}\Big(  \sum_{k \in{\cal P}} \frac{\mu_{1,k}^*\mathbf{H}_k}{\Gamma_{\mathrm{req}_1}}+\sum_{k\in {\cal B}} \frac{\beta_k^*\mathbf{H}_k}{\Gamma_{\mathrm{req}_1}}+\sum_{k\in {\cal I}}\delta_k^*\mathbf{H}_k\Big)\mathbf{W}^*_1.\notag
 \end{eqnarray}
Since $\mathbf{I}_{N_\mathrm{T}}+\phi^*\sum_{n=1}^{N_{\mathrm{T}}} \omega_n \mathbf{z}_n\mathbf{z}_n^H  $ is a positive definite matrix,  the following equality holds:
 \begin{eqnarray}
&&\hspace*{-6mm}\Rank\Big(\mathbf{W}^*_1\Big)=\Rank\Big(\Big(\mathbf{I}_{N_\mathrm{T}}+\phi^*\sum_{n=1}^{N_{\mathrm{T}}} \omega_n \mathbf{z}_n\mathbf{z}_n^H \Big)\mathbf{W}^*_1\Big)\notag \\
=&&\hspace*{-6mm}\Rank\Big(\Big(  \sum_{k \in{\cal P}} \frac{\mu_{1,k}^*\mathbf{H}_k}{\Gamma_{\mathrm{req}_1}}\hspace*{-1mm}+\hspace*{-1mm}\sum_{k\in {\cal B}} \frac{\beta_k^*\mathbf{H}_k}{\Gamma_{\mathrm{req}_1}}\hspace*{-1mm}+\hspace*{-1mm}\sum_{k\in {\cal I}}\delta_k^* \mathbf{H}_k\Big)\mathbf{W}^*_1\Big)\notag\\
\le&&\hspace*{-6mm} \Rank\Big(\sum_{k \in{\cal P}} \frac{\mu_{1,k}^*\mathbf{H}_k}{\Gamma_{\mathrm{req}_1}}\hspace*{-1mm}+\hspace*{-1mm}\sum_{k\in {\cal B}} \frac{\beta_k^*\mathbf{H}_k}{\Gamma_{\mathrm{req}_1}}\hspace*{-1mm}+\hspace*{-1mm}\sum_{k\in {\cal I}}\delta_k^* \mathbf{H}_k\Big)\le K.
\label{eqn:rank_option}
 \end{eqnarray}
 Thus, $\Rank\Big(\mathbf{W}^*_1\Big)\le \min\{K, N_{\mathrm{T}}\}$. On the other hand, a rank reduction approach can be used to show $\sum_{l=1}^L\Rank^2(\mathbf{W}^*_l)+\Rank^2(\mathbf{W}^*_{\mathrm{E}})\le  \card({\cal P}) L+\card({\cal B})+\card({\cal I})$ for $\phi=0$. Please refer to   \cite[Theorem 3.2]{JR:DP_rank} for a detailed proof.

\bibliographystyle{IEEEtran}
\bibliography{OFDMA-AF}

\end{document}